\documentstyle[12pt]{article}
\newcommand{\nn}{\nonumber}
\newcommand{\bd}{\begin{document}}
\newcommand{\ed}{\end{document}}
\newcommand{\bc}{\begin{center}} 
\newcommand{\ec}{\end{center}}
\newcommand{\be}{\begin{eqnarray}}
\newcommand{\ee}{\end{eqnarray}}
\newcommand{\eqn}{\global\def\theequation}
\newcommand{\sw}{sin^2 \theta_W}
\newcommand{\fbd}{f_B}
\renewcommand{\thefootnote}{\alph{footnote}}
\newcommand{\se}{\section}
\newcommand{\sse}{\subsection}
\newcommand{\bi}{\bibitem}
\begin{document}
\tolerance=10000
\baselineskip=7mm
\begin{titlepage}  

 \vskip 0.5in   
 \null
\begin{center}
 \vspace{.15in}
{\LARGE {\bf Radiative Leptonic B Decays in the Light Front Model
}
}\\
\vspace{1.0cm}
  \par
 \vskip 2.5em
 {\large
  \begin{tabular}[t]{c}
{\bf C.~Q.~Geng$^a$, C.~C.~Lih$^{a}$ and Wei-Min Zhang$^{a,b}$}
\\
\\
{\sl ${}^a$Department of Physics, National Tsing Hua University} 
\\  {\sl  $\ $ Hsinchu, Taiwan, Republic of China }\\
\\
and\\
\\ 
       {\sl ${}^b$Institute of Physics, Academia Sinica}\\
       {\sl  $\ $ Taipei, Taiwan,  Republic of China }\\
   \end{tabular}}
 \par \vskip 5.0em
 {\Large\bf Abstract}
\end{center}

Within the light front framework, we calculate the form factors for $B\to 
\gamma$ transitions directly in the entire physical range of momentum 
transfer. Using these form factors, we study the radiative decays of
$B\to l\nu_l\gamma$ and $B_{s(d)}\to \nu\bar{\nu}\gamma$.
We show that the decay rates of $B\to l\nu_l\gamma\ (l=e,\mu)$ and $B\to 
\nu\bar{\nu}\gamma$ are larger than that of the corresponding purely 
leptonic modes. Explicitly, in the standard model, we find that
the branching ratios of $B\to \mu\nu_{\mu}\gamma$ and $B_s\to 
\nu\bar{\nu}\gamma$ are $3.7\times 10^{-6}$ and $5.0\times 10^{-8}$,
in contrast with $3\times 10^{-7}$ and $0$ for $B\to \mu\nu_{\mu}$ and 
$B\to\nu\bar{\nu}$, respectively.

\end{titlepage}

\se{Introduction}
$\ \ \ $


It is known that the purely leptonic B decays of $B\to l\nu_l$ could be 
used to determine the weak mixing element of $|V_{ub}|$  
in the Cabibbo-Kobayashi-Maskawa matrix \cite{ckm} as well as the value of
the B meson decay constant $f_B$ \cite{r1}.
The decay rates of these purely leptonic modes are given by
\be
\Gamma (B\to l\bar{\nu }_{l})=\frac{G_{F}^{2}}{8\pi }%
|V_{ub}| ^{2}f_{B}^{2}\left( \frac{m_{l}^{2}}{m_{B}^{2}}\right)
m_{B}^{3}\left( 1-\frac{m_{l}^{2}}{m_{B}^{2}}\right) ^{2}\,.
\label{n1}
\ee
However, the rates for $B\to e\bar{\nu}_{e}$ and $\mu\bar{\nu}_{\mu}$ 
in Eq. (\ref{n1}) are helicity 
suppressed with the suppression factors of $m_l^2/m^2_B$ with  
$l=e$ and $\mu$, respectively, and one has that
$Br(B^-\to e^-\bar{\nu}_e,\mu^-\bar{\nu}_{\mu})\simeq (7\cdot 
10^{-12},3\cdot 10^{-7})$ by taking $|V_{ub}|=3\times 10^{-3}$, 
$f_B=200\ MeV$ and $\tau_{B^-}\simeq 1.65\ ps$ \cite{pdg}.
Clearly, it is difficult to measure these decays, especially for the 
light charged lepton mode.
Although there is no suppression for the $\tau$ channel, it is 
hard to observe the decay experimentally because of the low efficiency.
Similar helicity suppression effect is also expected in
the flavor changing neutral current (FCNC) processes of 
$B_{s(d)}\to l^+l^-$, which are sensitive to new physics beyond the 
standard model \cite{r2}.
Furthermore, to persevere the helicity conservation, the decays of 
$B_{s(d)}\to\nu\bar{\nu}$ are forbidden in the standard model.

Recently, there has been a considerable amount of theoretical 
attention 
\cite{rad1,rad2,rad3,rad4,rad6,rad7,rad8}
to the class of the radiative B decays, such as, $B\to l\nu_{l}\gamma$,
$B_{s(d)}\to l^+l^-\gamma$ and $B_{s(d)}\to\nu\bar{\nu}\gamma$.
These decays receive two types of contributions:
internal bremsstrahlung (IB) and
structure-dependent (SD) \cite{r8}.
The IB contributions are still helicity suppressed \cite{rad1}, 
while the SD ones contain the electro-magnetic coupling constant 
$\alpha$ but they are free of the helicity suppression. 
Therefore, the radiative decay rates of $B\to l_i\bar{l}_j\gamma$ 
($l_{i,j}=l,\nu_l$) could have an enhancement with respect to the purely 
leptonic modes of $B\to l_i\bar{l}_j$ due to the SD contributions.
Indeed, it has been shown that, for example, 
the branching ratios of $B\to \mu\nu_{\mu}\gamma$ 
\cite{rad1,rad2,rad3,rad4} and
$B_s\to \nu\bar{\nu}\gamma$ \cite{rad6,rad8} are $O(10^{-6})$ and 
$O(10^{-9})$, in contrast with that of $O(10^{-7})$ and $0$ for
the corresponding purely leptonic modes, respectively, in the standard model.
The measurements of the above decays in future B factories
provide an alternative way of knowing the B decay constants and the CKM 
matrix elements \cite{r1}.

In this paper, we concentrate on the radiative decays of $B\to l\nu_l
\gamma$ and $B\to \nu\bar{\nu}\gamma$.  We will use the light front 
formulation \cite{wmz,r10} to evaluate the hadronic matrix elements.
These decays have been studied in various quark models
\cite{rad1,rad2,rad3,rad4,rad6,rad7,rad8}.
It is known that as the recoil momentum increases, we have to
start considering relativistic effects seriously. In particular,
at the maximum recoil point, there is no reason to expect that the
non-relativistic quark model is still applicable. A consistent treatment 
of the relativistic effect of the quark motion and spin in a bound state 
is a main issue of the  relativistic quark model. The 
light front quark model \cite{lfm,hlfm} is the widely accepted relativistic 
quark model in which a consistent and relativistic treatment of quark 
spins and the center-of-mass motion can be carried out. In this paper 
we calculate the $P\to\gamma$ ($P$: pseudoscalar meson) form 
factors directly at time-like momentum transfers for the first time.
We will give their dependence on the momentum transfer $p^{2}$
in whole kinematic region of $0\leq p^{2}\leq p_{\max }^{2}$.

The paper is organized as follows. In Sec.~2, 
we present the relevant effective Hamiltonians for the radiative 
decays of $B\to l\bar{\nu}_l\gamma$ and $B_{s(d)}\to \nu\bar{\nu}\gamma$, 
respectively.
In Sec.~3, we study the form factors in the $B\to\gamma$ transition
within the light front framework.
We calculate the decay branching ratios
in Sec.~4.
We give our conclusions in Sec.~5.

\se{Effective Hamiltonian}

To study the decays of $B\to l\nu_l\gamma$, we start with the effective 
Hamiltonian for $b\to u\, l\nu_l$ at the quark level in the standard model, 
which is given by 
\be
H_{eff}(b\to u\, l\nu_l)= {G_F\over \sqrt{2}}V_{ub}\bar{u}
\gamma_{\mu}(1-\gamma_5)b\bar{\nu}\gamma_{\mu}(1-\gamma_5)l\,.
\label{he1}
\ee
For the radiative B decays, if we neglect the helicity suppressed photon 
emission from the final lepton, from Eq. (\ref{he1}) we get
\be
H_{eff}(B\to  l\nu_l\gamma)&=& {G_F\over \sqrt{2}}V_{ub}<\gamma|\bar{u}
\gamma_{\mu}(1-\gamma_5)b|B>\bar{\nu}_l\gamma_{\mu}(1-\gamma_5)l\,.
\label{He1}
\ee
For the processes of $B_{q}\to \nu_l\bar{\nu}_l\gamma\ 
(q=s,d;l=e,\mu,\tau)$, at the quark level, they arise from the box and 
$Z$-penguin diagrams, as shown in Fig. 1,
that contribute to $b\to q\nu_l\bar{\nu}_l$ with the photon 
emitting from the charged particles in the diagrams. However, when the 
photon line is attached to the internal charge lines as the W boson and 
t-quark lines, there is a suppression factor of $m_b^2/M_W^2$ in the 
Wilson coefficient in comparing with those in $b\to 
q\nu_l\bar{\nu}_l$ \cite{rad1}. 
Thus, we need only consider the diagrams with 
the photon from the external quarks. From the effective interactions for 
$b\to q\nu_l\bar{\nu}_l$, we obtain the effective Hamiltonians for 
$B_q\to\nu_l\bar{\nu}_l\gamma$ as follows:
\be
H_{eff}(B_q\to  \nu_l\bar{\nu}_l\gamma)=
{G_F\over \sqrt{2}}{\alpha\over   
2\pi\sin^2\theta_W}V_{tb}V_{tq}^*D(x_t)<\gamma|\bar{q}
\gamma_{\mu}(1-\gamma_5)b|B>\bar{\nu}_l\gamma_{\mu}(1-\gamma_5)\nu_l\,,
\label{He2}
\ee
where $x_t=m_t^2/M_W^2$ and 
\be
D(x_t)&=& {x_t\over 8}\left[-\,{2+x_t\over 1-x_t}+{3x_t-6\over (1-x_t)^2}\ln 
x_t\right]\,.
\label{Dxt}
\ee
We note that in Eqs. (\ref{He2}) and (\ref{Dxt}), 
only the leading contributions have been included and the additional
$1/m_b^2$ and $\alpha_s$ corrections to the result, which are small, can 
be found in Ref. \cite{mbalpha}.

\se{Form Factors on the Light Front}

 From the effective Hamiltonians in Eqs. (\ref{He1}) and (\ref{He2}), we 
see that to find the decay rates, we have to evaluate the hadronic matrix 
elements: $<\gamma|J_{\mu}|B>$, where 
$J_{\mu}=\bar{u}\gamma_{\mu}(1-\gamma_5)b$ with $u$ representing the 
light quarks of up, down and strange, respectively. The elements can be 
parameterized as follows:
\be
<\gamma (q)|\bar{u}\gamma^{\mu }\gamma_{5}b|B(p+q)>
&=& ie{F_{A}\over M_{B}}\left[
\epsilon ^{*\mu }( p\cdot q) -( \epsilon ^{*}\cdot p)
q^{\mu }\right]
\nn\\
<\gamma (q)| \bar{u}\gamma^{\mu}b|B(p+q)> &=&
ie{F_{V}\over M_{B}} \epsilon^{\mu \alpha \beta \gamma }\epsilon 
_{\alpha }^{*}p_{\beta }q_{\gamma } 
\label{n2}
\ee
where $q$ and $p+q$ are photon and $B$-meson four 
momenta, $F_{A}$ and $F_{V}$ are form factors of axial-vector and vector,
respectively, and $\epsilon $ is the photon polarization vector. 

The form factors in Eq. (\ref{n2}) will be calculated in the light front 
quark model at the time-like momentum transfers in which the physically 
accessible kinematic region is $0\leq p^{2}\leq p_{\max }^{2}$.
We consider that a meson bound state consists of a 
quark $q_{1}$ and an anti-quark $\bar{q}_{2}$ with 
total momentum $(p+q) $. For the $B$-meson bound state
we use the Gaussian-type wave function, given by \cite{r10,r14,r15}:
\be
	|B(p+q)>&=& \sum_{\lambda _{1}\lambda_{2}}\int [dk_{1}][dk_{2}] 
		2(2\pi)^{3}\delta ^{3}(p+q-k_{1}-k_{2}) \nn\\
	&& ~~~~~~~~ \times \Phi _{B}^{\lambda _{1}\lambda _{2}}(x,k_{\bot})
		b_{b}^{+}(k_{1},\lambda _{1}) d_{\overline{u}}^{+}(
		k_{2},\lambda _{2}) |0>\,,
\ee
where $k_{1(2)}$ is the on-mass shell light front momentum of the
internal quark $b(\bar{u})$.
The light front relative momentum variables $(x,k_{\bot })$
are defined by 
\be
	k_1^+= x(p+q)^{+}\,,\ k_{1\bot} = x(p+q)_{\bot}+k_{\bot}\,.
\ee

The normalization conditions can be written as
\be
	<B(p)|B(p')>=2(2\pi^{3}) p^{+}\delta ^{3}( p-p') \,,
\ee
which leads to 
\be
	\sum_{\lambda_1\lambda_2} \int \frac{dxd^{2}k_{\bot }}{2(2\pi)^{3}}
		| \Phi^{\lambda_1\lambda_2}_B (x,k_{\bot })| ^{2} =1\,.
		\label{n4}
\ee
The B meson wave function $\Phi_{B}^{\lambda _{1}\lambda _{2}}
\left( x,k_{\bot }\right) $ is chosen to be a Gaussian-type 
momentum distribution:
\be
	\Phi _{B}^{\lambda _{1}\lambda _{2}}(x,k_{\bot })=N\left( 
	\frac{2k_{1}^{+}k_{2}^{+}}{M_{0}^{2}-\left( m_{u}-m_{b}
	\right) ^{2}}\right)^{\frac{1}{2}}\overline{u}\left( k_{1},
	\lambda _{1}\right) \gamma^{5}v\left( k_{2},\lambda _{2}\right) 
	\sqrt{\frac{dk_{z}}{dx}} \exp \left( -\frac{\vec{k}^{2}}
	{2\omega_{B}^{2}}\right)   \, ,	\label{n6}
\ee
with
\be
	& & [dk_1]= {dk^+dk_{\bot}\over 2(2\pi)^3}\, , \ \
	N = 4 \left({\pi\over \omega_{B}^{2}}\right)^{3\over 4}
		\nn \\
	& & k_{z} =\left( x-\frac{1}{2}\right) 
	M_{0}+\frac{m_{b}^{2}-m_{u}^{2}}{2M_{0}} \, , \ \
	M_0^2={k^2_{\bot}+m_u^2\over x}+{k^2_{\bot}+m_b^2\over 1-x} \, ,
		\nn \\
	& & \sum_\lambda u(k,\lambda) \overline{u}(k,\lambda) =
		{m + \not \! k \over k^+} \, , \ \
	\sum_\lambda v(k,\lambda) \overline{v}(k,\lambda) =
		- {m - \not \! k \over k^+} \, , 
\ee
where the $\omega$ is a parameter related to the physical size of 
the meson, which is of order $\Lambda_{QCD}$. The value of $\omega$ 
ranges from $0.3$ to $0.6$ \cite{r12}. The spinors in Eq.(\ref{n6})
approximately take care the relativistic spin kinematics of quarks 
inside the B mesons.

The gauged photon state with momentum $p$ and spin $\lambda$ can be
described by:
\be
	|p\lambda>& =& N' \Bigg\{ a^{+}(p,\lambda) + \sum_{\lambda _{1},
		\lambda _{2}}\int [dk_{1}][dk_{2}] \Phi _{q\bar{q}}^{
		\lambda _{1}\lambda _{2}\lambda}(p,k_{1},k_{2})
			\nn\\
	&& ~~~~~~~~ \times 2(2\pi)^{3}\delta^{3}(p-k_{1}-k_{2}) b^{+}( k_{1},
		\lambda _{1}) d^{+} (k_{2},\lambda _{2}) \Bigg\} 
		| 0 > \, . \label{gbb}
\ee
The second term in Eq. ({\ref{gbb}) corresponds the photon state in QED 
in terms of quark pairs. Eq. (\ref{gbb}) satisfies the light-front 
bound state
\be
	H_{LF}|p,\lambda> ={p_{\bot}^{2}\over p^{+}}|p,\lambda >
		\label{n7}
\ee
with 
\be
	H_{LF}&=& H_{0}+H_{I}\,,
\ee
where $H_{0}$ is the free energy Hamiltonian of quarks and photons, 
and $H_{I}$ is the QED interacting part between quarks and photons
in the light-front gauge $A^+=0$, given by 
\be
	H_{I} &=& e_q \int q_{+}^{+}\{-2\frac{1}{\partial ^{+}}\partial^{i}
		A_{\bot}^{i}-\gamma \cdot A_{\bot }\frac{1}{\partial^{+}}
		(\gamma \cdot\partial _{\bot }-im)  \nn \\
	&& ~~~~~~~~~~ -\frac{1}{\partial ^{+}}(\gamma _{\bot }\cdot 
		\partial_{\bot}+im)\gamma \cdot A_{\bot }\}q_{+}{dx^{+}
		d^{2}k_{\bot } \over 2} \, , \label{n8}
\ee
and $e_q$ is the quarks' electric change, $q_+$ is the dynamical 
component of quark field on light-front: $q(x) = q_+(x) + q_-(x)$
with $q_\pm (x) = {1\over 2} \gamma^0\gamma^\pm q(x)$, and $A_\bot$
is the transverse component of the gauge field in the light-front
gauge. 

 From Eqs. (\ref{n7})-(\ref{n8}), we find the distribution of 
$\Phi_{q\bar{q}}^{\lambda _{1}\lambda _{2}\lambda }$ as
\be
	\Phi_{q\bar{q}}^{\lambda _{3}\lambda _{4}\lambda }(q,k_{1},k_{2}) 
		&=&\frac{e_q}{ED}\chi _{-\lambda _{2}}^{+}
		\left\{-2\frac{q_{\bot }\cdot \epsilon _{\bot }}
		{q^{+}}-\gamma _{\bot}\cdot \epsilon_{\bot }
		\frac{\gamma _{\bot }\cdot k_{2_{\bot }}-m_{2}}{k_{2}^{+}} 
		\right.		\nn \\
	&& ~~~~~~~ \left.-\frac{\gamma _{\bot }\cdot k_{1_{\bot }}-m_{1}}
		{k_{1}^{+}}\gamma_{\bot }\cdot \epsilon _{\bot }\right\}
		\chi _{\lambda_{1}}\,, \label{pqq}
\ee
with
\be
	ED&=&\frac{q_{\bot }^{2}}{q^{+}}-\frac{k_{1_{\bot}}^{2}+m_{1}^{2}}
		{k_{1}^{+}}-\frac{k_{2_{\bot }}^{2}+m_{2}^{2}}{k_{2}^{+}} \,.
\ee
Thus the gauge boson state wave function in Eq. (\ref{gbb}) can be rewritten
as
\be
	|\gamma (q)> &=& N' \Bigg\{a^{+}(q,\lambda) + \sum_{\lambda_{1}
		\lambda_{2}}\int [dk_{1}][dk_{2}]2(2\pi)^{3}\delta^{3}
		(q-k_{1}-k_{2}) \nn\\ 
	&& ~~~~~~~~~ \times \Phi _{q\bar{q}}^{\lambda_{1}\lambda_{2}\lambda}
		(q,k_{1},k_{2}) b_{q}^{+}(k_{1},\lambda _{1}) d_{\bar{q}}^{+} 
		(k_{1},\lambda _{2}) \Bigg\} | 0 > \, .
\ee

Since the transfer momenta in the decay processes are time-like, 
it is convenient to choose the light front coordinate: $p^{+}\geq 0$ 
and $p_\bot = 0$. By considering the ``+'' component in the weak 
current the matrix elements in Eq. (\ref{n2}) become
\be
	<\gamma (q)| u_{+}^{+}\gamma_{5}b_{+}|B(p+q)>
		&=&-ie\frac{F_{A}}{2M_{B}}\left( \epsilon
		_{\bot }^{*}\cdot q_{\bot }\right) p^{+}\,,
			\nonumber \\
	<\gamma (q)|u_+^+b_+|B(p+q)>&=&e\frac{F_{V}}{2M_{B}}
		\epsilon ^{ij}\epsilon _{i}^{*}q_{j}p^{+}\,.
			\label{ff}
\ee
The form factors of $F_{A}$ and $F_{V}$ in
Eq. (\ref{ff}) are found to be 
\be
	F_{A}(p^{2}) &=&i4M_{B}
		\int \frac{dx'd^{2}k_{\bot }}{2(2\pi)^{3}}\Phi
		\left( x,k_{\bot }^{2}\right) {x'-x\over x(1-x)}
			\nonumber \\
	&&~~~~~~~~ \times \left\{ \frac{1}{3}\frac{m_{b}+Bk_{\bot }^{2}
		\Theta}{m_{b}^{2}+
		k_{\bot}^{2}}-\frac{2}{3}\frac{m_{u}-Ak_{\bot }^{2}\Theta }
		{m_{u}^{2}+k_{\bot }^{2}}  \right\}\,, \label{fffa}
\ee
\be
	F_{V}(p^{2}) &=&i4M_{B}
		\int \frac{dx'd^{2}k_{\bot }}{2\left( 2\pi \right) ^{3}}\Phi
		\left( x,k_{\bot }^{2}\right) {x'-x\over x(1-x)}
			\nonumber \\
	&&\left\{ \frac{1}{3}\frac{m_{b}-(1-x)(m_{b}-m_{u}) k_{\bot }^{2}
		\Theta }{m_{b}^{2}+k_{\bot }^{2}}-\frac{2}{3}\frac{m_{u}-
		x\left( m_{b}-m_{u}\right) k_{\bot }^{2}\Theta }{m_{u}^{2}
		+k_{\bot }^{2}}\right\}\,, \label{fffv}
\ee
where
\be
	A &=& (1-2x') x(m_b-m_u) -2x'm_u\,, \nn\\
	B &=& 2(1-x') xm_b+(1-2x') (1-x)m_u\,,\nn\\
	\Phi (x,k_{\bot}^2) &=& N\left( 
		{2x(1-x) \over M_0^2-(m_u-m_b)^2}\right)^{1/2}
		\sqrt{{dk_{z}\over dx}}\exp \left( 
		-{\vec{k}^{2}\over 2\omega_B^2}\right)\,, \nn\\
	\Theta &=& {1\over \Phi(x,k_{\bot}^2) }
		{d\Phi(x,k_{\bot}^{2})\over dk_{\bot}^2} \, , \nn\\
	 x&=&x'\left(1-{p^2\over M_B^2}\right),\
		\vec{k}=(\vec{k}_{\bot},\vec{k}_{z}) \,.
\ee

To illustrate the form factors, we input the values of $m_{u}=0.3,$ 
$m_{b}=4.5,$ $M_{B}=5.2$, and $\omega =0.57$ in $GeV$ to integral 
whole range of $p^{2}$.  The results of $F_A$ and $F_V$ in the entire
range of momentum transfer $p^2$ are shown in Fig. 2.

\se{Decay Branching Ratios}
\sse{$B^{+}\to l^{+}\nu _{l}\gamma$}

For the radiative decays of $B^{+}\to l^{+}\nu _{l}\gamma$,
we will only consider the cases of $l=e$ and $\mu $. 
 From the effective Hamiltonian in Eq. (\ref{He1}) and the matrix element 
in Eq. (\ref{ff}), we find that
the amplitude of $B^{+}\to l^{+}\nu _{l}\gamma$ is
\be
M_{B^{+}\to l^{+}\nu \gamma } &=&-\frac{ieG_{F}V_{ub}}{\sqrt{2}}%
\epsilon _{\mu }^{*}H^{\mu \nu }\bar{u}\left( p_{\nu }\right) \gamma
_{\mu }\left( 1-\gamma _{5}\right) v\left( p_{l}\right)\,,
\ee
with
\be
H_{\mu \nu } &=&{F_A\over M_B}(-p\cdot q\,g_{\mu\nu}+p_{\mu }'
q_{\nu }) +i\epsilon _{\mu \nu \alpha \beta }\frac{F_{V}}{M_{B}}%
q^{\alpha }p'^{\beta}\,,  
\ee
where $p'$ and $q$ are $B$-meson and photon four momenta,
respectively, and $\epsilon _{\mu }$ is photon polarization vector. 
Since the form factors $F_{V,A}$ depend on the transfer momentum 
$p^{2}$, we need to replace $p^{2}$ into $(p',q)$. 
In the physical allowed region of $B^{+}\to l^{+}\nu _{l}\gamma $,
one has that
\be
m_{l}^{2}\leq p^{2}\leq M_{B}^{2}\,.
\ee
To describe the kinematic of the decay, two variables are needed. 
For convention, we defined 
$x''=2E_{\gamma}/M_B$ and $y=2E_l/M_B$ 
in the $B$-meson rest frame in order to easily write down 
momentum $p^{2}$ in term of $x''$, which has the form
\be
p^{2}&=& M_{B}^{2}(1-x'')\,.
\ee
We get the differential decay rate 
\be
\frac{d^{2}\Gamma ^{l}}{dx''d\lambda } 
&=&\frac{M_{B}}{256\pi
^{3}}\left| M\right| ^{2}=C\rho (x'',\lambda) ,
\ee
where $\lambda =(x''+y-1-r)/x''$,
\be
C &=&\frac{\alpha }{32\pi ^{2}}G_{F}^{2}M_{B}^{5}\left| V_{ub}\right| ^{2}\,,
\ee
and
\be
\rho (x,\lambda) &=&\rho_{+}(x'',\lambda) 
+\rho_{-}(x'',\lambda), 
\ee
with
\be
\nn
\rho_{+} &=&\frac{1}{2}|F_{A}+F_{V}|^{2}x''\lambda\left[(\lambda 
x''+r)(1-x'') -r\right] , 
\\
\nn
\rho_{-} &=&\frac{1}{2}|F_{A}-F_{V}|^{2}x''
(1-\lambda)\{(x''-1)[r+x''(\lambda -1)] +r\} , 
\\
r &=&\frac{m_{l}^{2}}{M_{B}^{2}}\,.
\ee

We write the physical region for $x''$ and $\lambda$:
as
\be
\nn
0 &\leq &x''\leq 1-r \,,
\\
\frac{r}{1-x''} &\leq &\lambda \leq 1\,.
\ee
In Fig. 3, we show the branching ratio of $B^+\to\mu^+\nu_{\mu}\gamma$ 
as a function of the parameter $\omega$, where
we have used $m_u=300\ MeV$, $|V_{ub}|\simeq 3\times 10^{-3}$ 
and $\tau_B\simeq 1.65\ ps$ \cite{pdg}.
For $\omega=0.57\ GeV$, 
we get the integrated 
branching ratios of $B^+\to l^+\nu_{l}\gamma$ as
\be
Br(B^{+}\to \mu^{+}\nu _{\mu}\gamma) &\simeq & 3.7\times 10^{-6}\,,
\\
Br(B^{+}\to e^{+}\nu_{e}\gamma) &\simeq & 3.5\times 10^{-6}\,.
\ee

\sse{$B_{s(d)}\to \nu \bar{\nu }\gamma$}

 From the effective Hamiltonians for 
$B_q\to\nu_l\bar{\nu}_l\gamma$ in Eq. (\ref{He2}) and
the form factors defined in Eq. (\ref{ff}), 
we can write the amplitude of $B_{q}\to \nu_l\bar{\nu}_{l}\gamma$ as
\be
M &=& -ie
{G_F\over \sqrt{2}}{\alpha\over  
2\pi\sin^2\theta_W}V_{tb}V_{tq}^*D(x_t)
\epsilon_{\mu }^{*}H^{\mu \nu }\bar{u}(p_{\bar{\nu}}) \gamma
_{\mu }(1-\gamma_{5})v(p_{\nu})\,,
\ee
with
\be
H_{\mu \nu } &=&{F_A\over M_B}(-p'\cdot q\,g_{\mu\nu}+p'_{\mu }q_{\nu })
+i\epsilon _{\mu \nu \alpha \beta }\frac{F_{V}}{M_B}q^{\alpha }p^{'\beta}\,. 
\ee
where the form factors are given by Eqs. (\ref{fffa}) and (\ref{fffv})
with the replacement of the light quark $(u)$ by $s$ and $d$ quarks, 
respectively.

Similar to the decays discussed in the previous subsection, 
we also define $x''=2E_{\gamma}/M_B$ and $y=2E_{\bar{\nu}}/M_B$ 
in the $B$-meson rest frame in order to re-scale the energies of the 
photon and anti-neutrino. 
By integrating the variable $y$ in the
phase space of variable $y$, we obtain the differential decay
rate of $B\to \nu\bar{\nu}\gamma$ as
\be
\frac{d\Gamma }{dx''} &=& 6\alpha \left({G_F\alpha\over 
16\pi^2\sin^2\theta_W}\right)^2
(|F_A|^2+|F_V|^2)|V_{tb}V_{tq}^*|^2D^2(x_t)x''^3(1-x'')
M_{B}^{5}\,,
\label{tdr}
\ee
where we have included the three generations of neutrinos.

Using $m_d=300\ MeV$, $m_s=400\ MeV$, $m_{t}=176\ GeV$, $|V_{tb}|=1$, 
$|V_{ts}|\simeq 0.04$, and $\omega=0.57$, the differential decay 
branching ratio $dBr(B_s\to \nu\bar{\nu }\gamma)/dx''$ as a function of 
$x''=2E_{\gamma}/M_B$ is shown in Fig. 4.
In Fig. 5, we give the branching ratio of $B_s\to \nu\bar{\nu }\gamma$ as 
a function of $m_t$. From the figure
we find that,
for $m_{t}=176\ GeV$ and $|V_{td}|\simeq 0.01$,
\be
\nn
Br(B_{s}\to \nu \bar{\nu}\gamma ) &=&5.0\times 10^{-8} \,,
\\
Br(B_{d}\to \nu \bar{\nu}\gamma ) &=&4.5\times 10^{-9}\,.
\ee

\se{Conclusions}

We have studied the form factors for $B\to \gamma$ transitions directly 
within the light front framework in the entire physical range of
momentum transfer. Using these form factors, 
we have calculated the radiative decays of
$B\to l\nu_l\gamma$ and $B_{s(d)}\to \nu\bar{\nu}\gamma$.
We have shown that the decays of $B\to l\nu_l\gamma\ (l=e,\mu)$ and 
$B\to \nu\bar{\nu}\gamma$ are dominated by the contributions from the 
diagrams with photon emission from the external quarks and thus 
overcome the helicity suppression effect.
We have found that, in the standard model, 
the branching ratios of $B\to e\nu_e\gamma$, $B\to \mu\nu_{\mu}\gamma$ and 
$B_{s(d)}\to \nu\bar{\nu}\gamma$ are $3.5\times 10^{-6}$, $3.7\times 
10^{-6}$ and $5.0\times 10^{-8}\ (4.5\times 10^{-9})$, respectively.
Some of the modes are clearly 
accessible in the future B factories.

\vspace{1cm}

\noindent
{\bf Acknowledgments}

We thank Professor A. Soni for useful discussions.
This work is supported by the National Science Council of the
ROC under contract numbers NSC86-2112-M-007-021, NSC86-2816-M001-009R-L,
and NCHC-86-02-007.

\newpage       
\begin{figure}[h]
\includegraphics{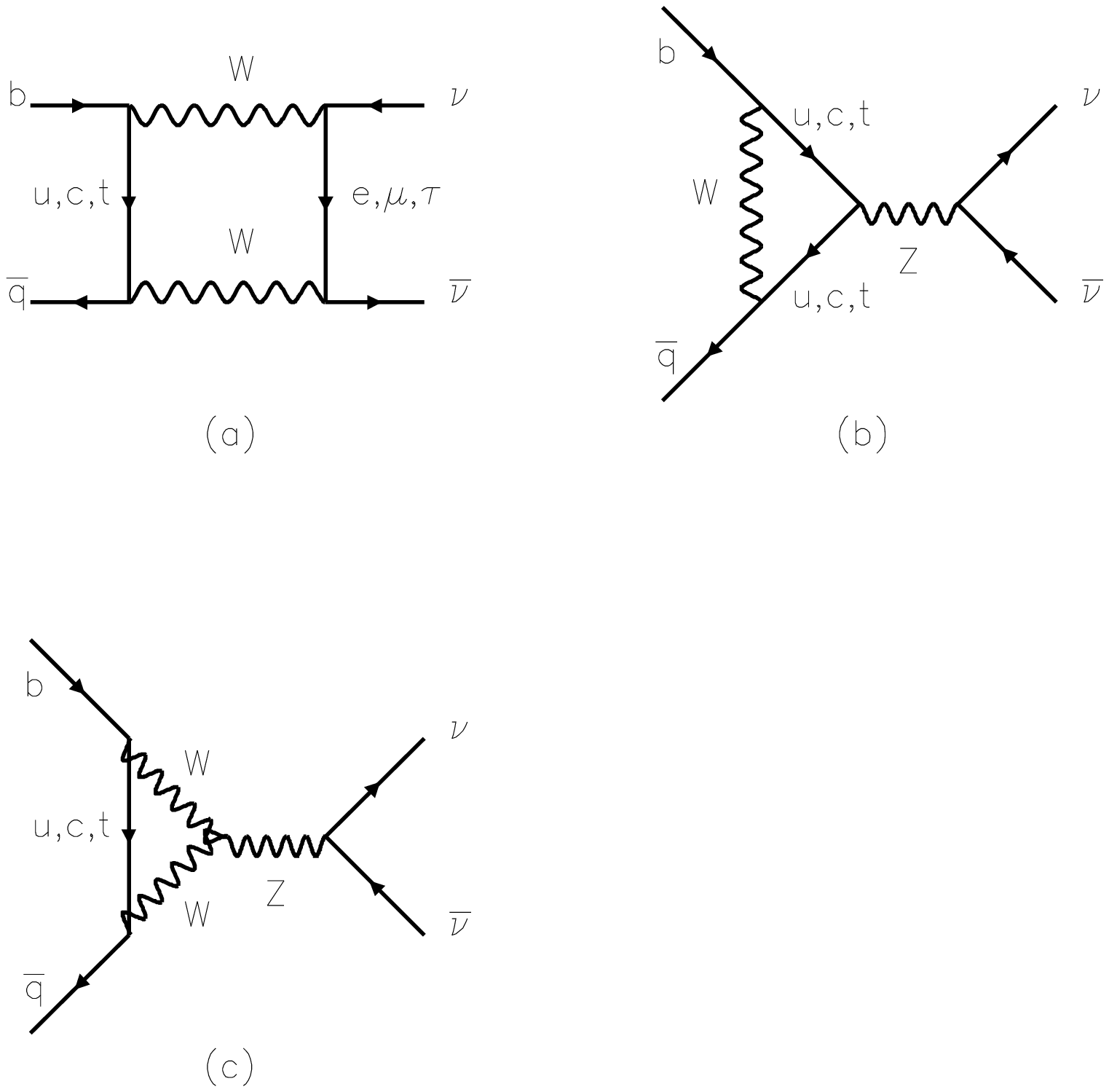}
\end{figure}
\vskip 10.cm
\bc
Fig. 1.
Loop diagrams that contribute $b\to q\nu\bar{\nu}$.
\ec

\newpage       
\begin{figure}[h]
\includegraphics{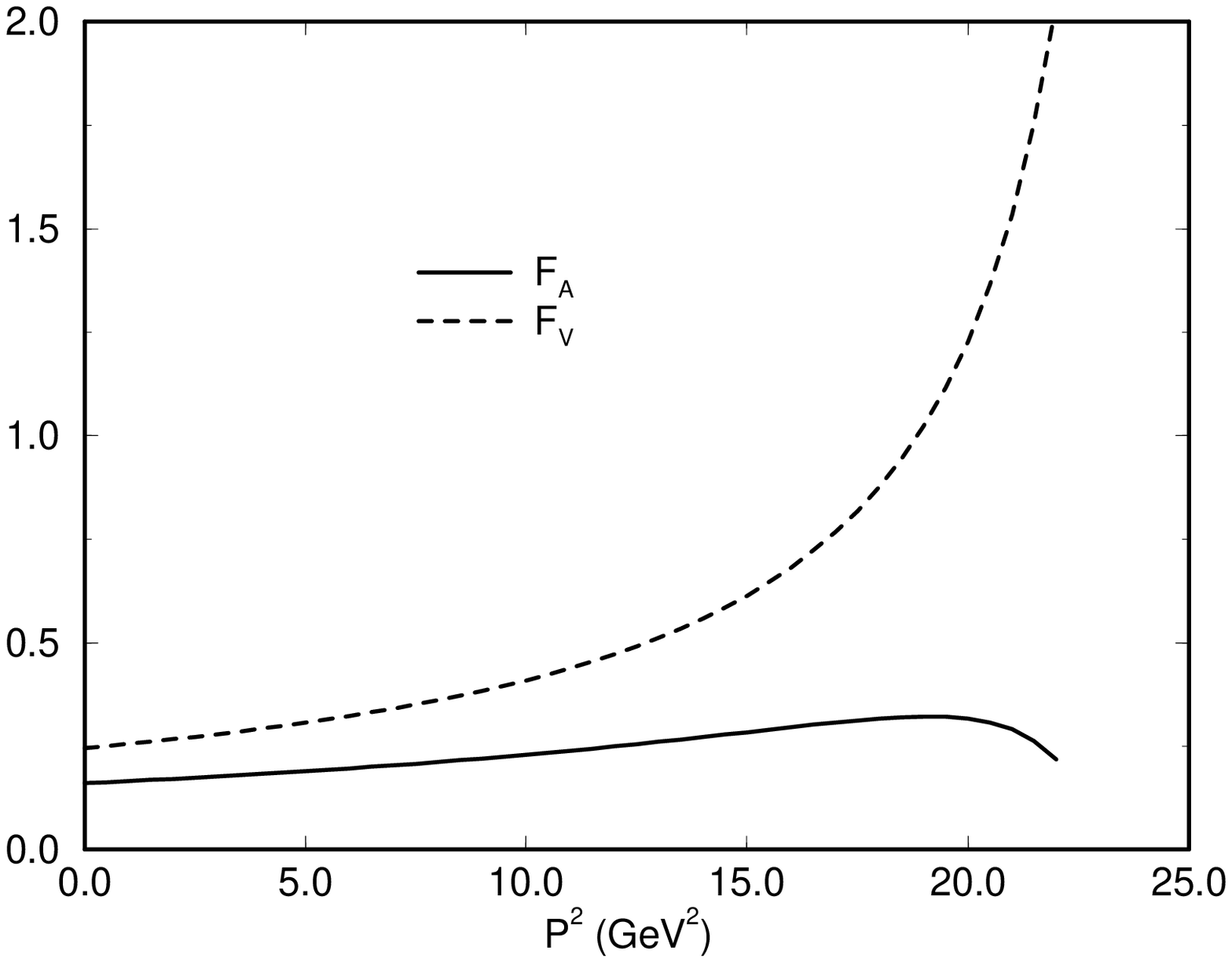}
\end{figure}
\vskip 8.cm
Fig. 2.
The values of the form factors $F_A$ (solid curve) and $F_V$ (dashed curve)
as functions of the momentum transfer $p^2$.

\begin{figure}[h]
\includegraphics{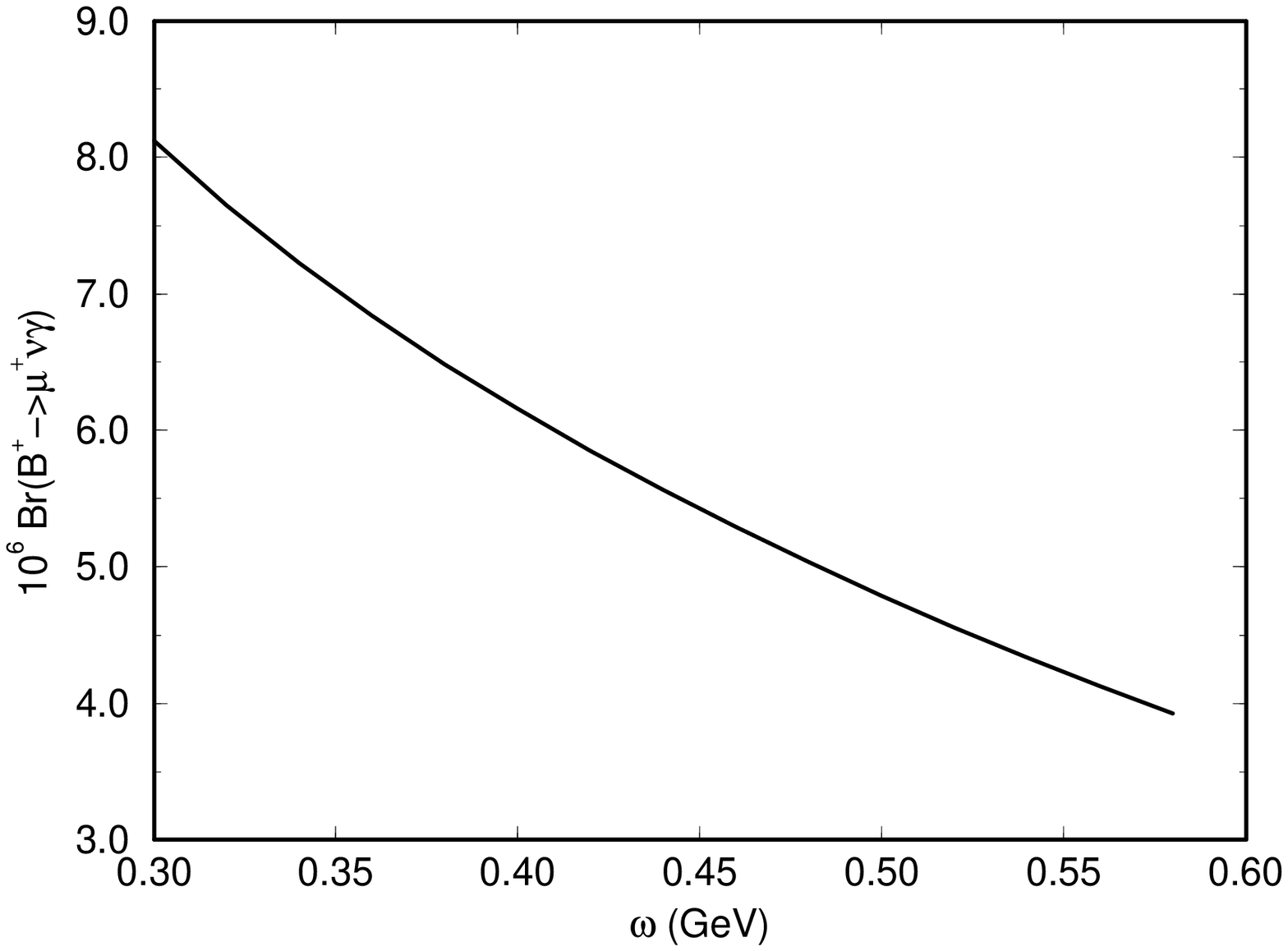}
\end{figure}
\vskip 9.cm
\bc
Fig. 3.
The branching ratio of $B^+\to\mu^+\nu_{\mu}\gamma$ as a function of
the parameter $\omega$.
\ec

\newpage       
\begin{figure}[h]
\includegraphics{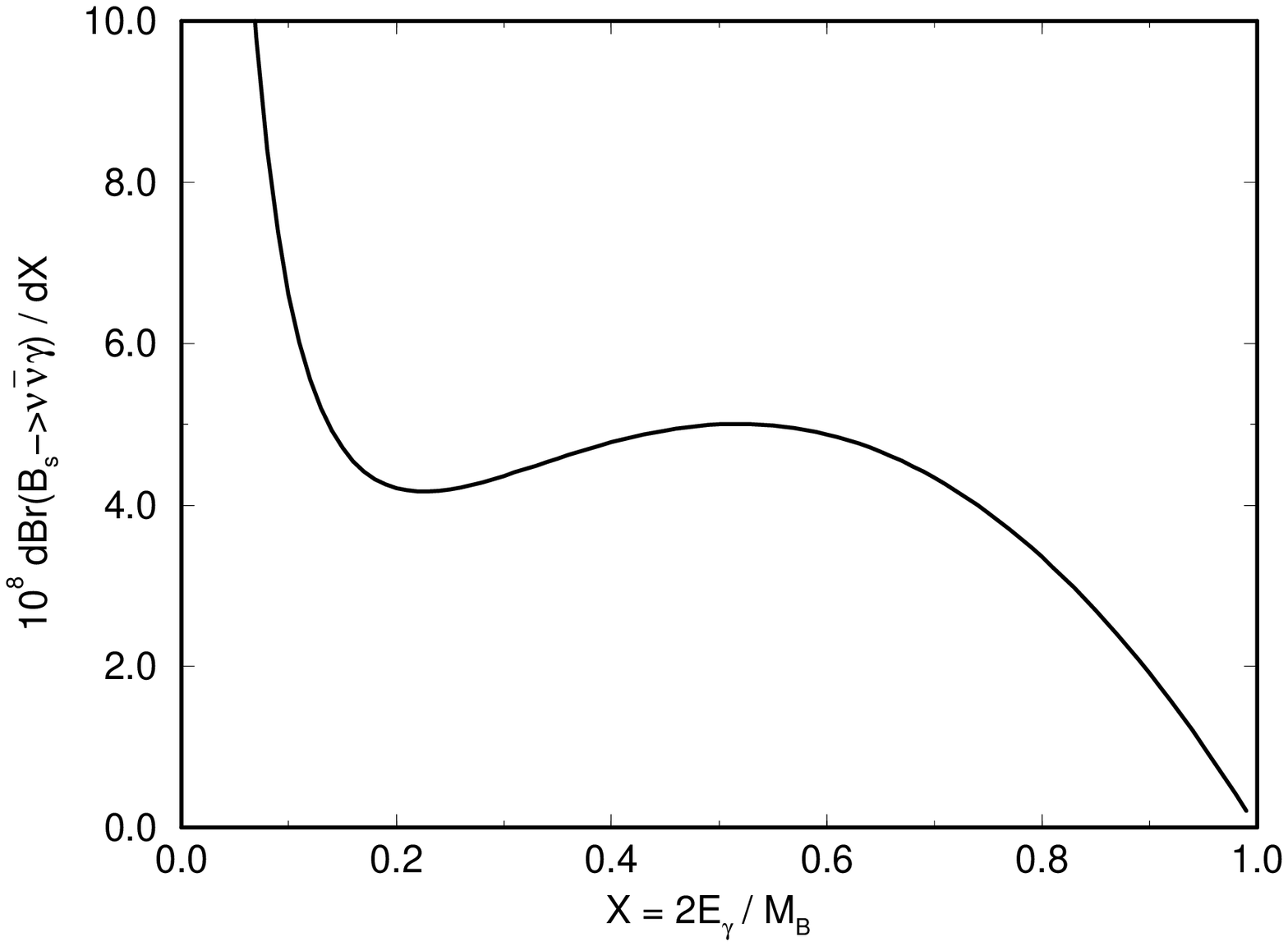}
\end{figure}
\vskip 8.cm
Fig. 4.
The differential decay 
branching ratio $dBr(B_s\to \nu\bar{\nu }\gamma)/dX$ as a function of 
$X=2E_{\gamma}/M_B$.

\begin{figure}[h]
\includegraphics{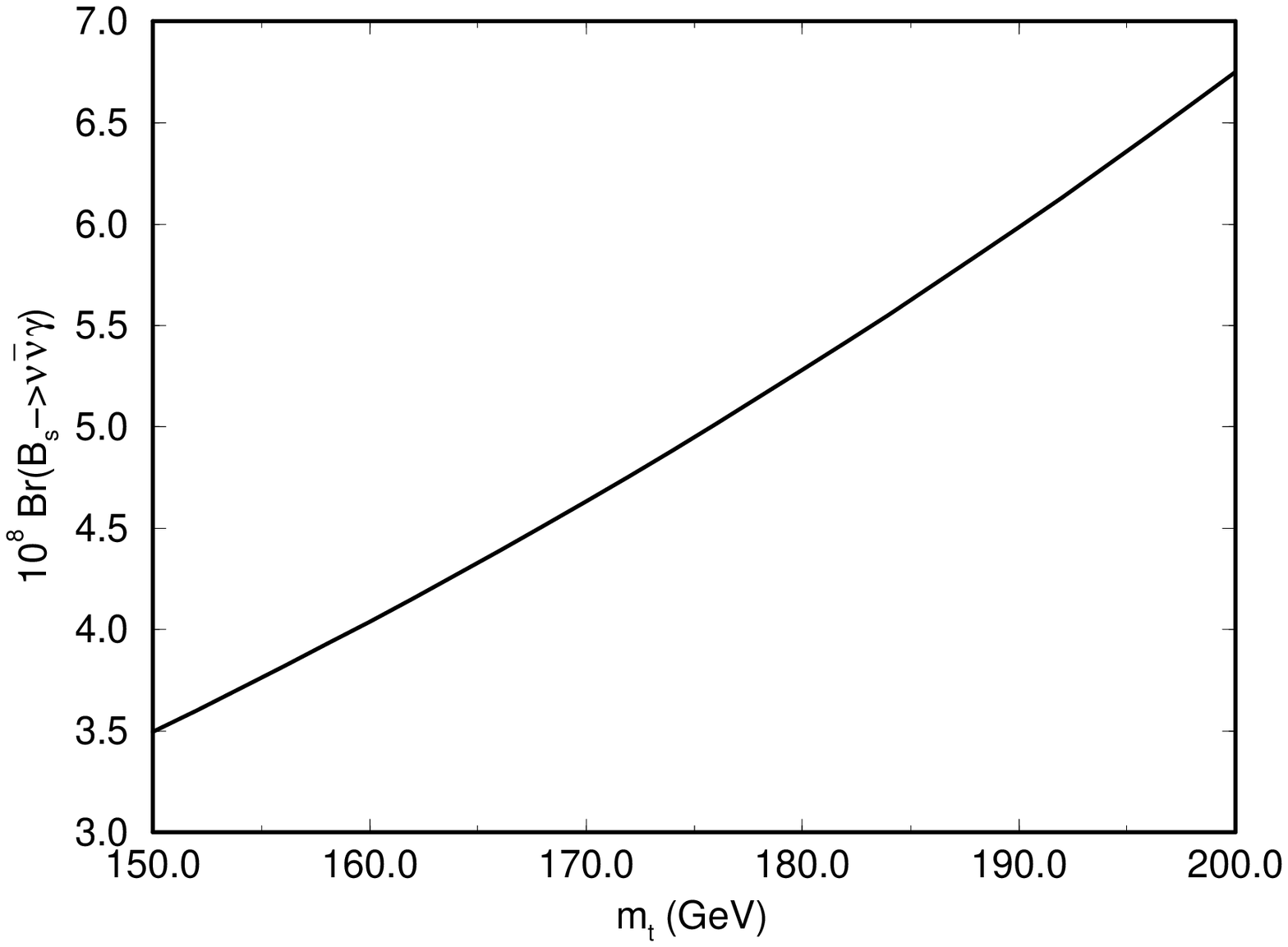}
\end{figure}
\vskip 9.cm
\bc
Fig. 5.
The branching ratio of $B_s\to \nu\bar{\nu }\gamma$ as 
a function of $m_t$. 
\ec

\ed

\end{document}